# Evaluation of a meta-analysis of the association between red and processed meat and selected human health effects

S. Stanley Young,[1] Warren Kindzierski[2]

[1]CGStat, Raleigh, NC 27607, USA; genetree@bellsouth.net
[2]12 Hart Place, St Albert, Alberta, Canada; wbk@shaw.ca

## Abstract

*Background*: Risk ratios or p-values from multiple, independent studies – observational or randomized – can be computationally combined to provide an overall assessment of a research question in meta-analysis. However, an irreproducibility crisis currently afflicts a wide range of scientific disciplines, including nutritional epidemiology. An evaluation was undertaken to assess the reliability of a meta-analysis examining the association between red and processed meat and selected human health effects (all-cause mortality, cardiovascular mortality, overall cancer mortality, breast cancer incidence, colorectal cancer incidence, type 2 diabetes incidence).

*Methods*: The number of statistical tests and models were counted in 15 randomly selected base papers (14%) from 105 used in the meta-analysis. Relative risk with 95% confidence limits for 125 risk results were converted to p-values and p-value plots were constructed to evaluate the effect heterogeneity of the p-values.

*Results*: The number of statistical tests possible in the 15 randomly selected base papers was large, median = 20,736 (interquartile range = 1,728–331,776). Each p-value plot for the six selected health effects showed either a random pattern (p-values > 0.05), or a two-component mixture with small p-values < 0.001 while other p-values appeared random. Given potentially large numbers of statistical tests conducted in the 15 selected base papers, questionable research practices cannot be ruled out as explanations for small p-values.

*Conclusions*: This independent analysis, which complements the findings of the original meta-analysis, finds that the base papers used in the red and resulting processed meat meta-analysis do not provide evidence for the claimed health effects.

# Introduction

The U.S. Food and Drug Administration (FDA) requires that evidence to support a health claim made for a foodstuff should be based on studies in humans (US FDA 2009). The randomized controlled trial (RCT), and in particular, randomized, placebo-controlled, double-blind intervention studies, provides the strongest evidence among studies in humans (Schneeman 2007). However, not all intervention studies on food and food components are RCTs, and frequently an RCT is unavailable and/or impractical.

In these cases, the FDA relies on lower-quality observational studies, especially cohort studies. These types of studies are dependent on dietary assessments using food frequency questionnaire (FFQ) analyses, which have proliferated in nutritional epidemiology (Byers 1999a, Freudenheim 1999, Sempos et al. 1999, Prentice 2010). Nutritional epidemiologists, other nutritional scientists, and food-policy analysts from the food industries, academia, and government all fund, approve, or conduct nutrition studies aimed at developing, supporting, and/or assessing food health claims (Byers 1999b). Nutritional epidemiology applies epidemiological methods to study a population about how diet affects health and disease in humans. Nutritional epidemiologists base most of their inferences about the role of diet (i.e., foods and nutrients) to cause or prevent chronic diseases on observational studies.

From the 1980s onward, the increase of computing capabilities facilitated the application of retrospective, self-administered dietary assessment instruments – the semi-quantitative food frequency questionnaire (FFQ) (Boeing 2013). A FFQ distributes a structured food list and a frequency response section to study participants, who indicate their usual frequency of intake of each food over a set period of time, usually a day or a week (Satija et al. 2015). FFQs, which are easy to use, place low burdens on participants, and aspire to capture long-term dietary intake. FFQs have become the most common method by which scientists measure intake in large observational study populations (Boing 2013, Satija et al, 2015).

A longstanding criticism of nutritional epidemiology to determine causality is that it relies predominantly on observational data, which researchers generally judge to be less reliable than experimental data (Satija et al, 2015). Causal criteria in nutritional epidemiology include (Potischman and Weed 1999): consistency, strength of association, dose response, plausibility, and temporality.

However, bias, which consists of systematic alteration of research findings due to factors related to study design, data acquisition, statistical analysis or reporting of results, can incapacitate any particular study's reliability as a way to apply these causal criteria (Boffetta et al. 2008, NASEM 2016 and 2019, Randall and Welser 2018). Selective reporting proliferates in published observational studies; researchers routinely test many questions and models during a study and then only report results that are interesting (i.e., statistically significant) (Gotzsche 2006).

Other acknowledged problems that limit the ability of nutritional epidemiology to substantiate claims of causal associations include (Byers 1999a):
- causal associations are difficult to prove in so complex a process as dietary intake, which includes interactions and synergies across different dietary components

- researcher flexibility allows estimates of food to be analyzed and presented in several ways – as individual food frequencies, food groups, nutrient indexes, and food-group-specific nutrient indexes
- researcher flexibility also allows dietary assessments to be presented with or without various adjustment factors, including other correlated foods and nutrients
- researcher flexibility allows scientists to choose among the many nutrient−disease hypotheses that could be tested
- classic criteria for causation are often not met by nutritional epidemiologic studies, in large part because many dietary factors are weak and do not show linear dose-response relationships with disease risk within the range of exposures common in the population

One aspect of replication – the performance of another study statistically confirming the same hypothesis or claim – is a cornerstone of science and replication of research findings is particularly important before causal inference can be made (Moonesinghe et al. 2007). However, an irreproducibility crisis currently afflicts a wide range of scientific disciplines including biomedical research (NASEM 2019).

For example, it is widely recognized that incomplete reporting is pervasive in biomedical research, leading to potential waste of resources, sceptical interpretation of findings and even scientific misconduct (Dickersin and Chalmers 2011). These types of situations far too frequently lead to the inability of scientists to replicate claims made in published research (Sarewitz 2012). Part of this replication problem may arise from researchers of observational studies performing large numbers of statistical tests and using multiple statistical models – referred to as MTMM (multiple testing and multiple modelling) (Westfall and Young 1993, Young and Kindzierski 2019).

Meta-analysis is a systematic procedure for statistically combining data from multiple studies that address a common research question, such as whether a particular food has an association with a disease (e.g., cancer) (Egger et al. 2001). Peace et al. (2018) recently evaluated 10 base papers included in a Malik et al. (2010) meta-analysis of the association between ingestion of sugar-sweetened beverages and risk of metabolic syndrome and type 2 diabetes. Peace et al. observed that the number of foods ranged from 60 to 165 across the 10 base papers, and that none of the base papers corrected for MTMM to account for chance findings. For the present evaluation we were interested in exploring whether the same sort of issues might be occurring with meta-analysis of the association between red and processed meats and selected human chronic effects.

## Methods

*Data Sets*
Vernooij et al. (2019) recently published a systematic review and meta-analysis of cohort studies pertaining to food health claims of red and processed meat. We selected six of 30 health effects that they reported on for further independent evaluation – specifically, those that had the largest number of base papers. These health effects included: all-cause mortality, cardiovascular mortality, overall cancer mortality, breast cancer incidence, colorectal cancer incidence, type 2 diabetes incidence. Upon our request, Vernooij et al. (2019) researchers provided data we used for this evaluation. We then used analysis search space counting (Peace et al. 2018, Young and Kindzierski 2019, Young et al. 2021) and p-value plots (Schweder and Spjøtvoll 1982) to assess the six health effect claims about red and processed meats.

The Vernooij et al. (2019) systematic review and meta-analysis reviewed 1,501 papers and selected 105 primary papers further analysis. The data set included 70 study cohorts. They used GRADE (*Grading of Recommendations Assessment, Development and Evaluation*) criteria (Guyatt et al. 2008) – which do not assess MTMM – to assess the reliability of the papers drawn from published literature and to select papers for their meta-analysis. Also, their study complied with recommendations of PRISMA (*Preferred Reporting Items for Systematic Reviews and Meta-Analyses*) (Moher et al. 2009).

Vernooij et al. (2019) stated that the base papers used in their meta-analysis, which were observational studies, provided low- or very-low-certainty evidence according to the GRADE criteria. Vernooij et al. concluded that, "*Low- or very-low-certainty evidence suggests that dietary patterns with less red and processed meat intake may result in very small reductions in adverse cardiometabolic and cancer outcomes*." In other words, their meta-analysis assigned little confidence to a claim that decreased consumption of red meat or processed meat improves health.

*Counting*
We were interested in counting three specific aspects related to the Vernooij et al. (2019) meta-analysis:

1. Initially, we want to present a general understanding of how common FFQ data is used by researchers investigating health effects in published literature. A potential problem is that researchers using FFQs – which are typically used in cohort studies – can subject their data sets to MTMM (Westfall and Young 1993) and produce large numbers of false positive results (Young et al. 2021). To understand how commonly FFQ data is used, we used a Google Scholar (GS) search of the literature to estimate the number of citations with the exact phrase "food frequency questionnaire" and a particular "health outcome".

    We chose 18 health effects for this search component, including: obesity, inflammation, depression, mental health, all-cause mortality, high blood pressure, lung and other cancers, metabolic disorders, low birth weight, pneumonia, autism, suicide, COPD (i.e., chronic obstructive pulmonary disease), ADHD (i.e., attention-deficit/hyperactivity disorder), miscarriage, atopic dermatitis, reproductive outcomes, erectile dysfunction.

2. Secondly, it is important to get a sense of the number of research questions under consideration in any given cohort study. It may be time-consuming and expensive to set up and follow a new cohort. But it can be relatively inexpensive to add new measurements and research questions to an existing cohort. For those reasons, it is typical to have many research questions under consideration with a given cohort study. Any single published study coming from a cohort study might appear only focused on one question. However, almost always there are many other questions at issue when you consider that the same cohort can be used many times over for research. Many papers based on a single cohort strongly imply there is no attention being paid to MTMM.

Scientists who conduct cohort studies generally use a straightforward statistical analysis strategy on the data they collect – e.g., what causes or risk factors are related to what outcomes (health effects). This strategy allows researchers to analyze large numbers of possible relationships. If a data set contains "C" causes and "O" outcomes, then scientists can investigate C x O possible relationships. They can also examine how an adjustment factor "A" (also called a covariate), such as parental age, income, education, or parity of child, can modify each of the C x O relationships.

We counted the number of questions considered in base papers that we randomly selected for review. For this we counted causes (C), outcomes (O), and adjustment factors (A); where the number of questions = $C \times O \times 2^A$.

3. Thirdly, we counted the number of published papers for each cohort study indicated in base papers that we randomly selected for review. We used a GS search to estimate the count (number of papers) that contain the data set name used by the cohort study. This count can be an over count, so for some data set names, we restricted the GS search of the name in the title of a paper.

*P-value Plots*
Epidemiologists traditionally use confidence intervals instead of p-values from a hypothesis test to demonstrate or interpret statistical significance. Since researchers construct both confidence intervals and p-values from the same data, one can be calculated from the other (Altman and Bland 2011a,b). We first calculated p-values from confidence intervals for all data used by Vernooij et al. (2019). We then developed p-value plots to inspect the distribution of the set of p-values after Schweder and Spjøtvoll (1982).

The p-value is a random variable derived from a distribution of the test statistic used to analyze data and to test a null hypothesis. In a well-designed study, the p-value is distributed uniformly over the interval 0 to 1 regardless of sample size under the null hypothesis and a distribution of true null hypothesis points plotted against their ranks in a p-value plot should form a straight line (Schweder and Spjøtvoll 1982, Hung et al. 1997, Bordewijk et al. 2020). Researchers can use the plot to assess the reliability of base papers used in meta-analyses.

We constructed and interpreted p-value plots as follows:
- We computed and ordered p-values from smallest to largest and plotted them against the integers, 1, 2, 3, …
- If the points on the plot followed an approximate 45-degree line, we concluded that the p-values resulted from a random (chance) process, and that the data therefore supported the null hypothesis of no significant association.
- If the points on the plot followed approximately a line with slope < 1, where most (the majority) of the p-values were small (less than 0.05), then the p-values provided evidence for a real (statistically significant) association.

- If the counts were high for a particular data set and points on a plot exhibited a bilinear shape (divided into two lines), then the p-values used for meta-analysis are consistent with a two-component mixture and a general (over-all) claim is not supported. In addition, the small p-value reported for the overall claim in the meta-analysis paper can not be taken as valid (Schweder and Spjøtvoll 1982).

P-value plotting is not itself a cure-all. P-value plotting cannot detect every form of possible systematic error. Questionable research procedures and publication bias may alter a p-value plot (Young et al. 2021). But this plotting procedure it is a useful tool allowing one to detect a strong likelihood that questionable research procedures – such as data reliability, p-hacking, HARKing – may have distorted base studies used in meta-analysis and rendered meta-analysis unreliable.

P-hacking involves the relentless search for statistical significance and comes in many forms, including multiple testing and multiple modelling without statistical correction (Ellenberg 2014, Hubbard 2015, Chambers 2017, Harris 2017, Streiner 2018). It enables researchers to find nominally statistically significant results even when there is no real effect; they convert a fluke, false positive into a statistically significant result (Boffetta et al. 2008, Ioannidis et al. 2011, McLaughlin and Tarone 2013, Simonsohn 2014). To HARK is to _hypothesize after the results are known_—to look at the data first and then come up with a hypothesis that has a statistically significant result (Randall and Welser 2018, Ritchie 2020).

P-value plotting is not the only means available by which to detect questionable research procedures. Other independent statistical tests may be available to account for frailties in base studies as they compute meta-analyses. Unfortunately, questionable research procedures in base studies severely degrade the utility of the existing means of detection (Carter 2019). We proffer p-value plotting as one possible approach to detect questionable research procedures in meta-analysis.

## Results

*Counting*
Table 1 shows how commonly FFQ data is used by researchers investigating various types of health outcomes in the Google Scholar literature for 18 health outcomes we selected (search performed 22 March 2021). Obesity associated with foods is a particular topic of interest with researchers. However, we can also see from Table 1 that health outcomes that we might expect to be less commonly associated with foods—such as reproductive outcomes and erectile dysfunction—are investigated cause-effect research topics.

A 5 to 20% sample from a population whose characteristics are known is considered acceptable for most research purposes as it provides an ability to generalize for the population (Creswell (2013). We accepted the Vernooij et al. (2019) judgment that their screening procedures selected 105 base papers with sufficiently consistent characteristics for use in meta-analysis. Based upon this, we randomly selected 15 of the Vernooij et al. (2019) 105 base papers (14%) for counting purposes. Table 2 shows count characteristics of 15 randomly selected papers from Vernooij et al. (2019).

**Table 1: Google Scholar search of health effects associated with foods (22 March 2021).**

| RowID | Outcome (effect) of interest | # of citations |
|---|---|---|
| 1 | obesity | 42,600 |
| 2 | inflammation | 23,100 |
| 3 | depression | 18,000 |
| 4 | mental health | 10,900 |
| 5 | all-cause mortality | 10,700 |
| 6 | high blood pressure | 9,470 |
| 7 | lung and other cancers | 7,180 |
| 8 | metabolic disorders | 5,480 |
| 9 | low birth weight | 4,630 |
| 10 | pneumonia | 2,140 |
| 11 | autism | 2,080 |
| 12 | suicide | 1,840 |
| 13 | COPD | 1,800 |
| 14 | ADHD | 1,370 |
| 15 | miscarriage | 1,240 |
| 16 | atopic dermatitis | 938 |
| 17 | reproductive outcomes | 537 |
| 18 | erectile dysfunction | 359 |

**Table 2. Characteristics of 15 randomly selected papers from Vernooij et al. (2019).**

| Citation# | Base Paper 1st Author | Year | Foods | Outcomes | Causes (Predictors) | Adjustment Factors (Covariates) | Tests | Models | Search Space |
|---|---|---|---|---|---|---|---|---|---|
| 8 | Dixon | 2004 | 51 | 3 | 51 | 17 | 153 | 131,072 | 20,054,016 |
| 31 | McNaughton | 2009 | 127 | 1 | 22 | 3 | 22 | 8 | 176 |
| 34 | Panagiotakos | 2009 | 156 | 3 | 15 | 11 | 45 | 2,048 | 92,160 |
| 38 | Héroux | 2010 | 18 | 32 | 18 | 9 | 576 | 512 | 294,912 |
| 47 | Akbaraly | 2013 | 127 | 5 | 4 | 5 | 20 | 32 | 640 |
| 48 | Chan | 2013 | 280 | 1 | 34 | 10 | 34 | 1,024 | 34,816 |
| 49 | Chen | 2013 | 39 | 4 | 12 | 5 | 48 | 32 | 1,536 |
| 53 | Maruyama | 2013 | 40 | 6 | 30 | 11 | 180 | 2,048 | 368,640 |
| 56 | George | 2014 | 122 | 3 | 20 | 13 | 60 | 8,192 | 491,520 |
| 57 | Kumagai | 2014 | 40 | 3 | 12 | 8 | 36 | 256 | 9,216 |
| 59 | Pastorino | 2016 | 45 | 1 | 10 | 6 | 10 | 64 | 640 |
| 65 | Lacoppidan | 2015 | 192 | 1 | 6 | 16 | 6 | 65,536 | 393,216 |
| 80 | Lv | 2017 | 12 | 3 | 27 | 8 | 81 | 256 | 20,736 |
| 92 | Chang-Claude | 2005 | 14 | 5 | 3 | 7 | 15 | 128 | 1,920 |
| 99 | Tonstad | 2013 | 130 | 1 | 4 | 10 | 4 | 1,024 | 4,096 |

Note: Citation# is Vernooij et al. reference number, Author name is first author listed for reference; Year = publication year; Foods = # of foods used in Food Frequency Questionnaire; Tests = Outcomes × Predictors; Models = $2^k$ where k = number of Covariates; Search Space = approximation of analysis search space = Tests × Models.

We note that while early food frequency questionnaire (FFQ) studies investigated only 61 foods (Willett et al. 1985), these 15 base studies include FFQ−cohort studies examining as many as 280 foods and 32 different health outcomes (Table 2). Summary statistics of the 15 base papers we randomly selected are presented in Table 3. We note that the median number of causes (predictors) was 15 and the median number of adjustment factors (covariates) was 9 in Table 3. These numbers by themselves suggest the great scope of the search space.

**Table 3. Summary statistics of 15 randomly selected papers from Vernooij et al. (2019).**

| Statistic | Foods | Outcomes | Causes (Predictors) | Adjustment Factors (Covariates) | Tests | Models | Search Space |
|---|---|---|---|---|---|---|---|
| minimum | 12 | 1 | 3 | 3 | 4 | 8 | 176 |
| lower quartile | 40 | 1 | 8 | 7 | 18 | 96 | 1,728 |
| median | 51 | 3 | 15 | 9 | 36 | 512 | 20,736 |
| upper quartile | 129 | 5 | 25 | 11 | 71 | 2,048 | 331,776 |
| maximum | 280 | 32 | 51 | 17 | 576 | 131,072 | 20,054,016 |
| mean | 93 | 5 | 18 | 9 | 86 | 14,149 | 1,451,216 |

Note: Foods = # of foods used in Food Frequency Questionnaire; Tests = Outcomes × Predictors; Models = $2^k$ where k = number of Covariates; Search Space = approximation of analysis search space = Tests × Models.

Nutritional epidemiologists may believe they gain an advantage by studying large numbers of outcomes, predictors, and covariates, on the presumption that this procedure maximizes their chances of discovering risk factor−health outcome associations (Willett et al. 1985). However, what they may have maximized is their likelihood of registering a false positive.

Given that the conventional threshold for statistical significance in most disciplines is a p-value of less than 0.05, a false positive result should occur 5% of the time by chance alone (Young et al. 2021). The median search space for the 15 randomly selected base papers was 20,736. We estimate that 5 percent of these 20,736 possible questions asked of a single typical FFQ−cohort data set underlying a nutritional epidemiology (observational) study will equal 1,037 chance findings that unwary researchers may mistake for a real result.

Finally, we wanted to offer a sense of the numbers of published papers for each cohort study that informed the 15 randomly sampled base papers as more evidence that MTMM is not considered. Table 4 presents cohort study names, an estimate of the number of papers in the Google Scholar literature for each cohort, and an estimate of the number of papers in the Google Scholar literature for each cohort using FFQs.

From Table 4 we observed that researchers conducted large quantities of statistical testing on the data from each cohort. We raise the potential concern that none of these 15 base papers provided correction for MTMM used on the same cohort−FFQ data set. Summary statistics for the 15 randomly sampled base papers from Table 4 are presented in Table 5. Based upon the count information presented, we observe that researcher using cohort study databases may examine large numbers of questions, in general and for FFQs, without proper attention to MTMM.

**Table 4. Cohort study names, an estimate of papers in literature for each cohort, and an estimate of papers in literature cohort using FFQs for the 15 randomly sampled base papers of Vernooij et al. (2019) (dd mmm yyyy).**

| Citation# | Author | Year | Cohort Study Name | Papers | Papers, Cohort+FFQ |
|---|---|---|---|---|---|
| 48 | Chan | 2013 | Mr. Os and Ms Os (Hong Kong) | 38,000 | 8 |
| 56 | George | 2014 | WHI Women's Health Initiative Observational Study | 37,200 | 1,520 |
| 49 | Chen | 2013 | HEALS and 'Bangladesh' | 12,400 | 1,080 |
| 53 | Maruyama | 2013 | JACC Japan Collaborative Cohort | 4,740 | 758 |
| 57 | Kumagai | 2014 | NHI Ohsaki National Health Insurance Cohort | 4,270 | 122 |
| 47 | Akbaraly | 2013 | Whitehall II study | 4,160 | 1,800 |
| 99 | Tonstad | 2013 | Adventist Health Study-2 | 2,630 | 653 |
| 80 | Lv | 2017 | China Kadoorie Biobank | 2,480 | 143 |
| 59 | Pastorino | 2016 | MRC National Survey of Health and Development | 1,860 | 148 |
| 31 | McNaughton | 2009 | Whitehall II study | 1,800 | 1,800 |
| 34 | Panagiotakos | 2009 | ATTICA Study | 1,650 | 1,650 |
| 8 | Dixon | 2004 | DIETSCAN (Dietary Patterns and Cancer Project) | 1,080 | 1,080 |
| 38 | Héroux | 2010 | ACLS (Aerobics Center Longitudinal Study) | 619 | 167 |
| 65 | Lacoppidan | 2015 | Diet, Cancer, and Health (DCH) cohort | 292 | 116 |
| 92 | Chang-Claude | 2005 | German vegetarian study | 18 | 13 |

Note: Citation# = Vernooij et al. (2019) reference number, Author name = first author listed for reference; Year = publication year; Cohort Name = name of study cohort; Papers = # of papers in literature mentioning study cohort; Papers, Cohort + FFQ = # of papers in literature mentioning study cohort using a Food Frequency Questionnaire (FFQ).

**Table 5. Summary statistics for estimate of papers in literature for the 15 randomly sampled base papers of Vernooij et al. (2019).**

| Statistic | Papers | Papers, Cohort+FFQ |
|---|---|---|
| minimum | 18 | 8 |
| lower quartile | 1,365 | 133 |
| median | 2,480 | 653 |
| upper quartile | 4,505 | 1,300 |
| maximum | 38,000 | 1,800 |
| mean | 7,547 | 737 |

Note: Papers = # of papers in literature mentioning study cohort; Papers, Cohort + FFQ = # of papers in literature mentioning study cohort using a Food Frequency Questionnaire (FFQ).

*P-value Plots*

The p-value plots for six health outcomes are presented in Figure 1. Each of the six images in Figure 1 indexes rank order (the x axis) and p-value (the y axis). The p-values – the dots in the body of the six images – are ordered from smallest to largest. The number of dots (p-values) in each image corresponds to the number of studies for each of the six health outcomes.

**Figure 1. P-value plots for meta-analysis of six health outcomes from Vernooij et al. (2019).**

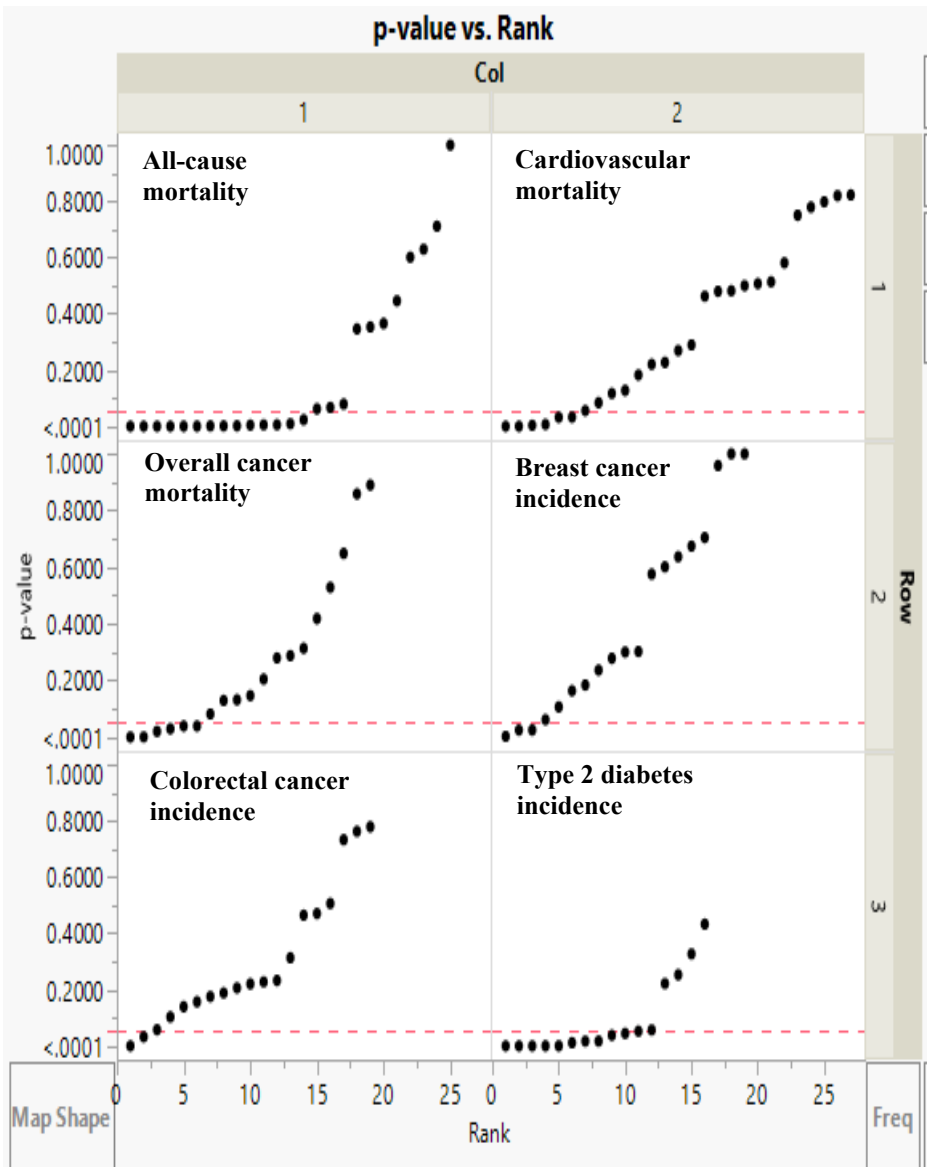

As noted in the Methods Section, if there is no effect the p-values will form roughly a 45-degree line. If the line is horizontal with most of the p-values small, then it supports a real effect. Finally, if the shape of the points is bilinear, then the results are ambiguous. The p-value plots for all-cause mortality, cardiovascular mortality, overall cancer mortality and type 2 diabetes incidence appear bilinear, hence ambiguous. The p-value plots for breast cancer incidence and colorectal cancer incidence appear as 45º lines, hence indicating a likelihood of no real association.

The image for colorectal cancer incidence is very unusual, with seven of the p-values on a roughly 45º line, two below the 0.05 threshold, and one extremely small p-value ($6.2 \times 10^{-5}$). Researchers usually take a p-value less than 0.001 as very strong evidence of a real effect (e.g., Boos and Stefanski 2011), although others suggest that very small p-values may indicate failures of research integrity (Al-Marzouki et al. 2005, Roberts et al. 2007, Bordewijk et al. 2020). If the small p-values indicates a real effect, then p-values larger than 0.05 should be rare.

The image for Type 2 diabetes incidence (lower right-hand side) has a p-value plot appearance of a real effect. On closer examination of the p-values and associated measured effects, however, the two smallest p-values ($4.1 \times 10^{-9}$ and $1.7 \times 10^{-7}$) have contrary results—the first is for a decrease of effect and the second is for an increase of effect. Our analysis might suggest some support for a real association between red or processed meat and Type 2 diabetes—but with the caution that the ambiguous results of the two smallest p-values makes us hesitant to endorse this result too strongly. We must note here a caution about failures of research integrity as suggested by others (Roberts et al. 2007, Redman 2013, Bordewijk et al. 2020).

Each health outcome presented in Figure 1 displays a wide range of p-value results – refer to Table 6. In the meta-analysis of breast cancer incidence (middle right-hand side image), for example, p-values ranged from < 0.005 to 1 across 19 base studies (> 2 orders of magnitude). In the meta-analysis of Type 2 diabetes incidence (lower right-hand side), the p-values ranged all the way from < $5 \times 10^{-9}$ to 1 (> 6 orders of magnitude). Such extraordinary ranges require a further caution about research integrity.

**Table 6. Minimum and maximum p-values for six health outcomes shown in Figure 1 from Vernooij et al. (2019).**

| Health outcome | Number of p-values | Minimum p-value | Maximum p-value |
|---|---|---|---|
| All-cause mortality | 25 | 1.61E-08 | 1 |
| Cardiovascular mortality | 27 | 6.43E-06 | 0.822757 |
| Overall cancer mortality | 19 | 0.000318 | 0.889961 |
| Breast cancer incidence | 19 | 0.002434 | 1 |
| Colorectal cancer incidence | 19 | $6.2 \times 10^{-5}$ | 0.779478 |
| Type 2 diabetes incidence | 16 | $4.1 \times 10^{-9}$ | 0.43304 |

The smallest p-value from Table 6 is $4.1 \times 10^{-9}$ – a value so small as to imply certainty (Boos and Stefanski 2011). A p-value this small may register a true finding – and small p-values are more likely in studies with large sample sizes (Young 2008). But the wide range of p-values in studies asking the exact same research question, however, including several studies which register results far weaker than p < 0.05, supports the idea that alternative explanations should be considered.

These possibly include some form of bias – systematic alteration of research findings due to factors related to study design, data acquisition, and/or analysis or reporting of results (Bofetta et al. 2008) and/or data fabrication (e.g., Mojon-Azzi and Mojon 2004a,b, Al-Marzouki et al. 2005, Eisenach 2009, Menezes et al. 2014, George and Buyse 2015, Dyer 2017, Klein 2017, Mascha et al. 2017, Miyakawa 2020).

Selective reporting is endemic in published observational studies where researchers routinely test many questions and models during a study and then only report/highlight statistically significant results (Gotzsche 2006). Given its potential importance, we further investigated circumstantial evidence of selective reporting for one of the six health outcomes presented in Figure 1 – all-cause mortality. To do this we conducted a GS search on 3 August 2021. We searched with the Cohort Study Name. The citation counts only approximate the number of papers/tests but point to extensive analysis of these cohorts.

Table 7 presents our results ranked by p-value, along with risk ratios (RRs) and confidence limits (CLs) from which we computed the p-values. From Table 7 it is inferred that a cohort study typically examines many outcomes, predictors, and covariates. The larger the number of citations, the greater the number of outcomes examined on a cohort study.

**Table 7. Characteristics of 24 cohort statistics for Vernooij et al. (2019) meta-analysis of all-cause mortality outcome.**

| Rank | Cohort Study Name | Citations | RR | $CL_{low}$ | $CL_{high}$ | p-value |
|---|---|---|---|---|---|---|
| 1 | Shanghai Men's Health Study | 1,390 | 1.49 | 1.33 | 1.67 | $1.61 \times 10^{-8}$ |
| 2 | Health Professionals' Follow-up Study | 86,400 | 1.31 | 1.19 | 1.44 | $1.17 \times 10^{-6}$ |
| 3 | Adventist Health Study | 6,270 | 1.25 | 1.15 | 1.36 | 3.06E-06 |
| 4 | Women's Health Initiative | 81,100 | 1.22 | 1.11 | 1.34 | 0.000177 |
| 5 | Nurses' Health Study | 125,000 | 1.17 | 1.08 | 1.26 | 0.000214 |
| 6 | Singapore Chinese Health Study | 5,580 | 1.14 | 1.06 | 1.23 | 0.001246 |
| 7 | Japan Public Health Center-based Prospective (JPHC) Cohort I [JPHC in title] | 420 | 0.91 | 0.85 | 0.96 | 0.00134 |
| 8 | Adventist Mortality Study | 270 | 1.20 | 1.09 | 1.34 | 0.001713 |
| 9 | Swedish Women's Lifestyle and Health cohort | 14,400 | 1.22 | 1.08 | 1.38 | 0.004045 |
| 10 | Black Women's Health Study | 2,160 | 1.33 | 1.12 | 1.58 | 0.004921 |
| 11 | Shanghai Women's Health Study | 5,440 | 1.15 | 1.05 | 1.26 | 0.00511 |
| 12 | Health Food Shoppers Study | 85 | 0.90 | 0.83 | 0.98 | 0.008966 |
| 13 | Adventist Health Study 2 (AHS-2) M | 2,670 | 1.11 | 1.02 | 1.21 | 0.02324 |
| 14 | Adventist Health Study 2 (AHS-2) W | 2,670 | 1.10 | 1.00 | 1.21 | 0.061948 |
| 15 | Third National Health and Nutrition Examination Survey Men | 136,000 | 1.43 | 1.04 | 1.96 | 0.066926 |
| 16 | Seguimiento Universidad de Navarra (SUN) project | 1,430 | 1.52 | 1.04 | 2.20 | 0.078877 |
| 17 | European Prospective Investigation into Cancer and Nutrition (EPIC) | 37,500 | 1.37 | 0.80 | 2.34 | 0.346286 |
| 18 | German Vegetarian Study | 18 | 0.91 | 0.74 | 1.12 | 0.353189 |
| 19 | Aerobics Center Longitudinal Study | 3,690 | 1.18 | 0.86 | 1.64 | 0.365669 |
| 20 | Third National Health and Nutrition Examination Survey | 219,000 | 1.14 | 0.83 | 1.55 | 0.445927 |
| 21 | PREvencion con Dieta MEDiterranea trial (PREDIMED) | 3,600 | 1.04 | 0.74 | 1.04 | 0.601206 |
| 22 | Health, Aging, and Body Composition Study (Health ABC) | 23,000 | 1.08 | 0.80 | 1.45 | 0.629478 |
| 23 | Whitehall II study | 31,400 | 1.05 | 0.82 | 1.35 | 0.711523 |
| 24 | Oxford Vegetarian Study | 587 | 1.00 | 0.87 | 1.15 | 1 |

Note: There are 25 studies and 24 cohorts as two studies used the same cohort; Rank = p-value rank; Citations = # of citations in literature mentioning study cohort; RR = Relative Risk; CL$_{low}$ = lower confidence limit; CL$_{high}$ = upper confidence limit.

## Discussion

Many nutritional epidemiologists now believe that red and processed meat are associated with severe health effects (e.g., Battaglia Richi et al. 2015 and Ekmekcioglu et al. 2018). The International Agency for Research on Cancer (IARC), the cancer agency of the World Health Organization, has classified red meat as probably carcinogenic to humans and processed meat as certainly carcinogenic to humans (WHO 2015). Our findings suggest that this consensus is problematic.

Other have challenged this consensus on other grounds. For example, Vernooij et al. (2019) suggest that the base observational studies are unreliable. Delgado et al. (2021) state that epidemiological studies reveal no authoritative connections between the intake of red and processed meat and occurrence of cardiovascular disease. Even the popular press has offered differing opinions – not least by citing popular low-carbohydrate and high-meat diets (Atkins, etc.) that do not seem to have imposed ill effects on their practitioners (Bueno 2013, Castellana 2021, Taubes 2021).

Vernooij et al. (2019) has offered plausible scientific arguments against red and processed meat─health effect claims. Their large-scale systematic review and meta-analysis of the 105 base papers studying the health effects of red and processed meats has provided evidence that the base papers, generally observational studies, offer low- or very-low-certainty evidence according to GRADE criteria. Many nutritional epidemiologists reacted to their research negatively. Some asked the editor of *Annals of Internal Medicine*, which accepted their study, to withdraw the paper before publication (Monaco 2019, Arends 2020).

Here we presented further supportive evidence, arrived at by an independent line of critique, that the studies that claiming health effects for red and processed meats may be unreliable. Our study suggests that the base studies, properly examined statistically (counting and p-value plots) for false positives and possible research integrity violations, do not support the claims.

Performing large numbers of statistical tests without offering all findings (which is now possible with supplemental material and web posting) makes it challenging to ascertain how many true or false-positive versus negative findings might exist in an analysis. Nutritional epidemiologists' failure to correct for multiple testing may register epidemiologists' larger failing.

Stroup et al. (2000) provided a proposal for reporting meta-analysis of observational studies in epidemiology. This proposal is frequently referred to in published literature (15,612 Google Scholar citations as of 1 May 2021). However, Stroup et al. make no mention of observational studies' MTMM problem and offer no recommendation to control for MTMM.

Meta-analyses may provide greater evidentiary value if and only if they combine results from base studies that: 1) use reliable data and analysis procedures, and 2) crucially all studies are responding to the same process (Fisher 1950, DerSimonian and Laird 1986). Base studies that do not correct for MTMM do not provide reliable data for meta-analyses. Furthermore, meta-analyses that combine base studies that do and do not correct for MTMM are not combining comparable studies. Either flaw suffices to render ineffective any meta-analysis that relies on even a single base study that fails to correct for MTMM.

Our bilinear p-value plots in Figure 1 suggest evidence that nutritional epidemiological meta-analyses have combined base studies that do not use comparable methods. Alternately, the bilinear plots may register the existence of one or more powerful covariates correlated with a predictor variable in some of the studies – that, for example, cardiorespiratory fitness was confounded with dietary risk of mortality (Héroux et al. 2010). The existence of an unrecognized covariate would also render irrelevant the meta-analysis' results.

The exceedingly large analysis search spaces in the 15 randomly selected base papers of Vernooij et al. (2019) – refer to Tables 2 and 3 – also make it plausible to infer that the small p-values among the base studies may be derived from some questionable research procedure such as p-hacking, which other researchers have shown is widespread (Head et al. 2015). The large number of papers derived from these cohort studies in Tables 4 and 5 strengthens this proposition.

Epidemiology studies that test many null hypotheses tend to provide results of limited quality for each association due to limited exposure assessment and inadequate information on potential confounders (Savitz and Olshan 1995). Epidemiologists tend to seek out small but (nominally) significant risk factor–health outcome associations (i.e., those that are less than 0.05) in multiple testing environments (Kindzierski et al. 2021). These studies also tend to generate more errors of false-positive or false-negative associations (Rothman 1990).

These practices may render their research susceptible to reporting false-positives as real results, and to risk mistaking an improperly controlled covariate for a positive association. A set of studies in a meta-analysis where number of statistical tests is large and whose p-values demonstrate bilinearity in a p-value plot should properly be regarded as suspect. The independent statistical methods that we employed in this evaluation – simple counting and p-value plots – suggest that the findings of the Vernooij et al. (2019) meta-analysis is reliable.

## Findings

Our independent evaluation – simple counting and p-value plots – suggests that the base papers used in the Vernooij et al. (2019) red and processed meat meta-analysis do not support evidence for the claimed health effects. This finding is consistent with that of Vernooij et al. (2019).